\documentclass{article}
\usepackage{graphicx}

\usepackage[margin=1in]{geometry}
\usepackage[utf8]{inputenc}
\usepackage{microtype}
\usepackage{amsmath,amsthm,amssymb, color, tikz, mathtools}
\usetikzlibrary{arrows.meta}
\usepackage{caption}
\usepackage{hyperref} 
\hypersetup{colorlinks,linkcolor={blue},citecolor={blue},urlcolor={blue}}

\newtheorem{theorem}{Theorem}
\newtheorem{corollary}[theorem]{Corollary}
\newtheorem{lemma}[theorem]{Lemma}

\newtheorem{observation}[theorem]{Observation}

\usetikzlibrary{positioning}

\newtheorem{definition}[theorem]{Definition}

\title{A King for Almost Everyone}
\date{}

\author{Ziad Ismaili Alaoui \thanks{University of Liverpool, United Kingdom. \texttt{\{ziad.ismaili-alaoui, nikhil.mande\}@liverpool.ac.uk}}\;\;\thanks{Philipps-Universität Marburg, Germany.}  \and Oded Lachish \thanks{University of London, Birkbeck, United Kingdom. \texttt{o.lachish@bbk.ac.uk} \\ \hspace*{1.8em}Ziad Ismaili Alaoui was funded by EPSRC grant EP/X039447/1.}\and Nikhil S. Mande \footnotemark[1]}

\begin{document}

\maketitle

\begin{abstract}
\noindent A tournament is an orientation of a complete graph. A vertex $v$ \emph{controls} a vertex $u\neq v$ if there is a directed path of length at most two from $v$ to $u$, and a $\delta$-king is one that controls at least $\delta(n-1)$ vertices. In the query complexity model, each query reveals the orientation of an edge. Shen, Sheng and Wu [SICOMP'03] established an $\Omega(n^{4/3})$ lower bound on the deterministic query complexity of finding a $1$-king, and a non-matching upper bound of $O(n^{3/2})$. Neither bound has been improved in over two decades, despite substantial progress on variants of the problem.

For $\delta$-kings with $\delta < 1$, Biswas, Jayapaul, Raman and Satti [Discrete Applied Mathematics'22] showed that $O(n^{4/3})$ queries suffice to find a $\frac12$-king. Lachish, Reidl and Trehan [FSTTCS'22] improved this guarantee to $\frac12+\frac{2}{17}$ using $\widetilde{O}(n^{4/3})$ queries, and asked whether a larger fraction could be achieved within the same query bound.

We answer this question affirmatively for every fixed fraction below one: for every fixed $\varepsilon\in(0,1)$, we give a deterministic algorithm that finds a $(1-\varepsilon)$-king using $O(n^{4/3})$ queries. 

The design of our algorithm is informed by a structural analysis combining a sparse regularity lemma due to Allen, Kohayakawa, Mota and Parente [Electronic Journal of Combinatorics'14] with a new sampling argument based on a reachability theorem of Afek, Gafni and Linial [arXiv'19].
\end{abstract}

\tableofcontents

\newpage

\section{Introduction}
\label{s:intro}
A \emph{tournament} is an orientation of a complete graph where, between every pair of distinct vertices $u,v$, there is exactly one arc, either $u\to v$ or $v\to u$.\footnote{In this paper, we refer to directed edges as ``arcs''.}  Landau~\cite{Lan53} in the 50s, who was motivated by dominance relations among chickens pecking each other, studied how to identify a ``most-dominant chicken'' when the pecking order is represented by a tournament. To do this, he introduced the notion of a \emph{king} (which was later coined by Maurer~\cite{M80}): A vertex from which every other vertex is reachable by a directed path of length at most two. So if we interpreted an arc $u \to v$ as $u$ \emph{defeating} $v$, a king is a vertex that defeats every other vertex either directly or through an intermediary vertex. This idea generalises the notion of a source, which has an outgoing arc to every other vertex. Obviously, not every tournament has a source, but it always has a king; indeed, it turns out that every vertex of maximum out-degree is a king~\cite{Lan53,M80}. Kings can be seen as a natural notion of a winner when pairwise comparisons are not necessarily transitive.  And more generally, tournaments and their properties have been studied extensively in graph theory and social choice theory; for more information, see Moon's monograph~\cite{Moon15} and the references in~\cite{MD24}. The existence of a king raises a natural algorithmic question:\vspace{1em}

\centerline{\textit{How much of a tournament must be revealed before one can find a king?}}\vspace{1em}

\noindent We study this question in the model of \emph{query complexity}. For an unknown tournament $T$ on $n$ vertices (the number of vertices is always known in advance), a query algorithm can only access the tournament's arcs by querying an oracle for two vertices $u,v \in [n]$ of its choosing. The oracle only returns whether $u \to v$ in $T$, or $v \to u$. The cost (i.e.,~query complexity) of an algorithm is the largest number of queries made over all inputs of size $n$. 

Shen, Sheng and Wu~\cite{SSW03} (also see~\cite{AFHN15}) gave an $O(n^{3/2})$-query deterministic algorithm for finding a king in an $n$-vertex tournament and an adversary establishing an $\Omega(n^{4/3})$ lower bound. These bounds have remained unchanged for over two decades. A stronger analysis of the Shen-Sheng-Wu adversary cannot help strengthen the lower bound, as Biswas, Jayapaul, Raman and Satti~\cite{BJRS22} later gave an algorithm that finds a king against it using $O(n^{4/3})$ queries. However, if we allow randomisation, we get a substantially smaller worst-case query cost of $\Theta(n)$~\cite{MPSS24, AGNS24, MPS23}.

\subsection{Approximate kings and our results}

A natural relaxation is to seek a vertex that reaches not all but many other vertices within two arcs. We say that a vertex $v$ \emph{controls} a vertex $u\neq v$ if there is a directed path of length at most two from $v$ to $u$. We also say that $u$ \emph{beats} or \emph{dominates} $v$ when $u \to v$ (note that $u$ may control $v$ but not beat it). For $\delta\in[0,1]$, a \emph{$\delta$-king} is a vertex that controls at least $\delta(n-1)$ vertices. In particular, a $1$-king is a king. We then ask:\vspace{1em}

\centerline{\textit{How much of a tournament must be revealed before one can find a vertex}}\centerline{\textit{controlling a fraction $\delta$ of the other vertices?}}\vspace{1em}

\noindent Biswas, Jayapaul, Raman and Satti~\cite{BJRS22} showed that $O(n^{4/3})$ queries suffice to find a $\frac12$-king. Lachish, Reidl and Trehan~\cite{LRT22} later improved the guarantee to $\frac12+\frac{2}{17}$ using $\widetilde O(n^{4/3})$\footnote{The tilde hides polylogarithmic factors; more formally, $f(n) \in \widetilde O(g(n))$ and $f(n) \in O(g(n)\text{polylog}(n))$ are equivalent.} queries. They explicitly asked whether a vertex controlling a larger fraction of the tournament could be found deterministically within the same asymptotic query bound~\cite[Section~5]{LRT22}.

In this work, we answer this question affirmatively for every fixed fraction below one; that is:

\begin{theorem}[Main]\label{thm:main}
For every fixed $\varepsilon\in(0,1)$, there is a deterministic algorithm that finds a $(1-\varepsilon)$-king in an $n$-vertex tournament using $O_\varepsilon(n^{4/3})$ queries.
\end{theorem}

\noindent Here, we use the subscript $\varepsilon$ for complexity bounds that depend on a function of $\varepsilon$; that is, $f(n) \in O_\varepsilon(g(n))$ implies there exists a function $h$ such that $f(n) \in O(g(n)h(\varepsilon))$. This also holds for $\Omega$ and $\Theta$. 

The analysis of Ajtai, Feldman, Hassidim and Nelson~\cite{AFHN15}, restated
in~\cite[Lemma~14]{BJRS22}, shows that the adversary
of Shen, Sheng and Wu~\cite{SSW03} requires
$\Omega(t^{4/3})$ queries to find a vertex controlling
at least $t$ other vertices.
Taking $t=\lceil(1-\varepsilon)(n-1)\rceil$ gives
a matching $\Omega_\varepsilon(n^{4/3})$ lower bound.
Thus, our result determines the optimal deterministic query
complexity of finding a $(1-\varepsilon)$-king,
up to a constant factor depending on $\varepsilon$,
for every fixed $\varepsilon\in(0,1)$.

\subsection{Overview of our approach}

Our algorithm makes two sequential batches of queries,
inspired by a similar approach used by Lachish, Reidl and Trehan~\cite{LRT22}. First, we fix a sparse
undirected graph on $n$ vertices with at most $n^{4/3}$
edges, which we call the \emph{template graph}.
Its key property is that the edge density between any
two disjoint sets, each containing at least a suitably
chosen constant fraction of the vertices, is close to
$p=n^{-2/3}$. We use a probabilistic argument to show that such
a graph exists for all sufficiently large $n$.

Our main contribution is a structural argument showing
that, whatever orientations the first batch reveals,
we can select $O_\varepsilon(n^{2/3})$ vertices whose
induced subtournament is guaranteed to reveal a
$(1-\varepsilon)$-king. The two batches work as follows.

\begin{enumerate}
\item \textit{First batch: the template graph's edges.}
We query the orientation of every template edge in
the unknown tournament. A sparse directed version
of Szemerédi's regularity lemma due to Allen,
Kohayakawa, Mota and Parente~\cite[Theorem~2.1]{AKMP14}
allows us to partition the vertices into a bounded
number of equal-sized groups, together with a small
exceptional group (the number of groups
depends only on $\varepsilon$) such that almost every pair of non-exceptional groups is
\emph{regular} in both directions; that is, in either direction, the arc density between
sufficiently large subsets is approximately the same
as the arc density between the entire groups in that
direction, with a small additive error compared with $p$. The template graph's property
also ensures that the majority direction between
two groups has density at least a constant multiple
of $p$.

We record these directions in an oriented graph $R$
whose vertices represent the non-exceptional groups.
Whenever both ordered pairs $(V_i,V_j)$ and $(V_j,V_i)$
are regular, we include the arc $i\to j$ if more arcs
go from $V_i$ to $V_j$ than in the reverse direction,
breaking ties by choosing $i\to j$ when $i<j$.
Since only a small proportion of pairs fail regularity
in at least one direction, $R$ is almost a tournament.

    \item \textit{Second batch: the union of the courts.}
    We carefully choose a small subset of each group, called its
    \emph{court}. An inheritance theorem for sparse
    lower-regularity due to Gerke, Kohayakawa, Rödl and
    Steger~\cite[Theorem~3.6]{GKRS07} allows us to choose
    courts of size $\Theta_\varepsilon(n^{2/3})$ with
    the following property: Whenever $i\to j$ is an arc
    of $R$, every sufficiently large subset of court
    $i$ sends queried arcs (from the first batch) to almost all vertices of
    group $j$. Here, ``sufficiently large'' means at
    least a fixed fraction of the court.

    These courts are chosen using only the first batch
    of answers. We then query all edges in the subtournament induced by
    their union. A sampling argument based on a
    two-tournament theorem of Afek, Gafni and
    Linial~\cite[Theorem~1]{AGL19} shows that, whatever
    these new answers are, some vertex $w$ beats a
    sufficiently large subset of each court in a
    collection whose outgoing arcs in $R$ cover almost
    all groups.

    The court property now gives us the required paths:
$w$ has a direct arc to each vertex in the selected
subsets of the courts, and these subsets collectively
send template arcs to almost all vertices of the
covered groups. Each such target is therefore reachable
from $w$ within two queried arcs. By choosing the parameters
so that at most $\varepsilon(n-1)$ other vertices remain
uncovered, we make $w$ a $(1-\varepsilon)$-king.
\end{enumerate}

\noindent The analysis of the number of queries made is easy:
The first batch uses at most $n^{4/3}$ queries.
Each court has $\Theta_\varepsilon(n^{2/3})$ vertices,
and there are $O_\varepsilon(1)$ groups, so their union
$Q$ has $\Theta_\varepsilon(n^{2/3})$ vertices.
Revealing its induced subtournament therefore uses
at most $O_\varepsilon(n^{4/3})$ additional queries.

It may be tempting to improve this bound by making the
template graph sparser. The known lower bound already
rules out an $o_\varepsilon(n^{4/3})$-query algorithm,
but our construction also illustrates the trade-off
directly. For a template with edge density $\Theta(p)$,
the first batch uses $\Theta(n^2p)$ queries.
The inheritance theorem allows us to choose courts
of size $L=\Theta_\varepsilon(1/p)$, giving
$|Q|=\Theta_\varepsilon(1/p)$. Revealing all pairs
within $Q$ brings the total cost to
$\Theta_\varepsilon(n^2p+p^{-2})$. Thus, within our construction, making the template
sparser increases the court size and the cost of
revealing their union. The two terms balance at
$p=n^{-2/3}$, which explains our choice of density and
gives the optimal query bound of
$\Theta_\varepsilon(n^{4/3})$.

\subsection{Comparison with related work}

The most relevant works to ours are those of Biswas, Jayapaul, Raman and Satti~\cite{BJRS22}, and of Lachish, Reidl, and Trehan~\cite{LRT22}.

Biswas et al.~gave a very simple $O(n^{4/3})$-query algorithm to find a $1/2$-king. Here is the idea, ignoring rounding and leftover vertices. Partition the vertices into blocks of size $k=n^{1/3}$, each with a representative that beats every other vertex in its block. To build each block, query all edges among $2k$ remaining vertices: some vertex beats at least $k-1$ others, so take it together with $k-1$ of its out-neighbours as a block and remove them. This costs $O(k^2)$ queries per block, hence $O(nk)=O(n^{4/3})$ queries overall. Now query all edges among the $n/k=n^{2/3}$ representatives, using another $O(n^{4/3})$ queries. A simple counting argument shows that some representative beats at least half the other representatives, and therefore reaches every vertex in their blocks within two steps. Together with the vertices it beats in its own block, this gives at least half of all other vertices.

Lachish et al.~improved the fraction to $1/2+2/17$ using $\widetilde{O}(n^{4/3})$ queries. Their algorithm introduces the template-graph approach that also underlies our result: First query a fixed sparse graph, then use its orientation to choose a small set of vertices for further queries. The properties of their template graph, however, are considerably different from ours: They require at least one edge between every two disjoint sets of size at least $\kappa n^{2/3}$. We consider much larger sets of size at least $\kappa n$, but require the number of edges between them to lie between $(1-\kappa)n^{-2/3}|X||Y|$ and $(1+\kappa)n^{-2/3}|X||Y|$. In a nutshell, their condition guarantees connectivity between smaller sets, while ours controls edge densities between larger sets.

Standard probabilistic arguments establish the existence of both types of template graphs. The second phase of Lachish et al.'s algorithm looks for a vertex that beats many vertices in carefully chosen small sets. These out-neighbours then serve as stepping stones to reach a large fraction of the tournament. One easy case is when they find a small set such that every subset containing at least half its vertices collectively dominates enough vertices. Querying all edges within this set then suffices: a maximum-outdegree vertex provides the required stepping stones.

When no such set exists, they partition almost all vertices into small sets called weak tiles, each with a small out-neighbourhood in the queried template. Each tile therefore has many vertices outside it that it does not beat. By the template property, any sufficiently large subset of these outside vertices collectively beats almost all of the tile. Their task is then to find a vertex with enough out-neighbours among these outside vertices for sufficiently many tiles. They achieve this through a sequence of selections and queries involving small vertex sets. They also explain~\cite[Section 4]{LRT22} how this second phase can actually be implemented using a single batch of queries to all edges in a subtournament.

Thus, both algorithms first query a sparse template and then query the subtournament induced by a small vertex set chosen from the revealed orientation. The difference lies in the structural argument used to choose this set. Lachish et al.~\cite{LRT22} use a dichotomy between ultra sets and weak tilings, followed by an analysis of the free sets associated with the tiles. A bottleneck in this analysis is the number of tiles for which they can guarantee coverage. In the proof of their Theorem~22, the tiles are partitioned into two families $\mathcal W_1$ and $\mathcal W_2$. Claims~23 and~24 guarantee that the chosen vertex reaches almost all vertices in every tile of $\mathcal W_1$, but give this guarantee for only approximately a $(1/2-\delta)$-fraction of the tiles of $\mathcal W_2$, where $\delta>0$ is their target improvement over $1/2$. This fraction remains bounded away from one as the template error parameter tends to zero, and the tiles in $\mathcal W_2$ can together contain a constant fraction of all vertices. Thus, these bounds alone do not extend to a $(1-\varepsilon)$-king for arbitrarily small $\varepsilon$. We instead use sparse regularity and an inheritance lemma to construct courts, and a sampling lemma to show that querying their union suffices to find a $(1-\varepsilon)$-king. Extending their result to this guarantee therefore calls for a structural argument that overcomes the loss in tile coverage.

\subsection{Related notions and other models}

Kings are closely connected to tournament solutions in
social choice theory. Maiti and Dey~\cite{MD24} studied the query complexity of computing several such solutions, including the Copeland set, the Slater set, the Markov set, the bipartisan set, the uncovered set, and the Banks set.
The uncovered set consists precisely of all kings, so
computing this set is a different task from finding a
single king. Since every tournament has a king, finding
one is a total search problem, a connection explored
in~\cite{GGLS26}. Related work has also studied the
complexity of recognising kings in succinctly represented
tournaments~\cite{HHTW10}.

The concept of kings also arises in the context of unreliable comparisons.
Ajtai, Feldman, Hassidim and Nelson~\cite{AFHN15} studied maximum finding with imprecise comparisons: elements have unknown numerical values, and comparisons may be adversarial when the values differ by at most a fixed threshold $\Delta$. Directing each comparison from the reported winner to the loser, deterministic maximum finding with additive error $2\Delta$ is easily seen to be equivalent to finding a king.

Other models have also been studied. The randomized and quantum query complexities of finding a king are now understood~\cite{MPS23, MPSS24, AGNS24}. Mande, Paraashar,
Sanyal and Saurabh~\cite{MPSS24} studied kings in the context of communication complexity, and we~\cite{alaoui2026directedcutquerycomplexityfinding} recently studied it from the perspective of directed-cut query complexity, where a query returns the number of arcs leaving
a specified vertex set.

Abboud, Grossman, Naor and Solomon~\cite{AGNS24} studied
finding multiple kings, when they exist. Ismaili Alaoui and Mande~\cite{IAM25}
studied the complexity of finding strong kings, a variant of kings. Weimann and Yuster~\cite{weimann2026maintaining} studied kings in the context of dynamic algorithms.
Finally, Afek, Gafni and Linial~\cite{AGL19} considered
communication over tournaments that change between
rounds. They proved that, for any two tournaments
$T_1,T_2$ on the same vertex set, some vertex reaches
every other vertex by an arc in either tournament,
or by an arc in $T_1$ followed by an arc in $T_2$.
This two-tournament theorem is a key ingredient
in our sampling argument.

\subsection{Outline}

The paper is organised as follows. Section~\ref{s:prelims} introduces the notation and terminology used throughout the paper. Section~\ref{s:properties} defines the template graph and states the structural lemmas underlying our approach. In Section~\ref{s:algorithm}, we use these lemmas to describe the algorithm and establish its correctness and query complexity. Section~\ref{s:structural} provides the proofs of the structural lemmas. Finally, we conclude in Section~\ref{s:conclusion} with a discussion of open questions and an AI statement.

\section{Preliminaries}
\label{s:prelims}
For a positive integer $n \geq 1$, we write $[n] \coloneq \{1, 2, \dots, n\}$.

\begin{definition}[tournament, subgraph, subtournament]
    A \emph{tournament} $T = (V, E)$ is a \emph{complete} oriented graph where $V$ is the set of vertices, $E$ is the set of arcs, and for every arc $(u, v) \in E$, $(v, u) \notin E$. For a directed or undirected graph $G=(V,E)$ and
    $V'\subseteq V$, the \emph{induced subgraph} $G[V']$
    has vertex set $V'$ and contains precisely the arcs
    or edges of $G$ whose endpoints both belong to $V'$.
    An induced subgraph of a tournament is called a
    \emph{subtournament}.
\end{definition}

\noindent All graphs considered are finite and have no loops or parallel edges. We use $V(G)$, $E(G)$ and $G[X]$ for the vertex set, edge set and subgraph induced by $X \subseteq V(G)$, respectively. An \emph{arc} is oriented; we always refer to an \emph{edge} as undirected.

\begin{definition}[$\delta$-king]
\label{def:king}
A $\delta$-\emph{king} in an oriented $n$-vertex graph is a vertex $v$ such that, for at least $\delta(n-1)$ other vertices $u \neq v$, at least one of the following holds:
\begin{itemize}
    \item $(v,u)$ is an arc in the graph, or
    \item there exists a vertex $w$ such that $(v,w)$ and $(w,u)$ are arcs in the graph.
\end{itemize} 
\end{definition}

\noindent For disjoint vertex sets $X,Y\subseteq V(G)$ in an undirected graph $G$, let $E_G(X,Y)$ denote the set of edges between them. For an oriented graph $D$ and disjoint sets $X,Y\subseteq V(D)$, let $E_D(X,Y)$ denote the set of arcs directed from $X$ to $Y$, and write $d_D^+(X,Y):=|E_D(X,Y)|$. When both sets are non-empty, define $\rho_G(X,Y):=|E_G(X,Y)|/(|X||Y|)$ and $\rho_D(X,Y):=|E_D(X,Y)|/(|X||Y|)$.

For a vertex $v$ of an oriented graph $D$, its out-neighbourhood is $N_D^+(v):=\{u:(v,u)\in E(D)\}$ and its out-degree is $d_D^+(v):=|N_D^+(v)|$. For a vertex set $X\subseteq V(D)$, write $N_D^+(X):=\bigcup_{v\in X}N_D^+(v)$; this union may contain vertices of $X$. We write $N_D^{++}(v):=(N_D^+(v)\cup N_D^+(N_D^+(v)))\setminus\{v\}$ for the set of other vertices reachable from $v$ by a directed path of length at most two. A set $X$ \emph{dominates} a set $Y$ in $D$ if $Y\subseteq N_D^+(X)$.

Let $D$ be an oriented graph, and let $\alpha\in(0,1)$ and $p\in(0,1]$. An ordered pair $(X,Y)$ of disjoint non-empty vertex sets is \emph{$(\alpha,p)$-regular} if $|\rho_D(X',Y')-\rho_D(X,Y)|\le\alpha p$ whenever $X'\subseteq X$ and $Y'\subseteq Y$ satisfy $|X'| > \alpha|X|$ and $|Y'| > \alpha|Y|$. This condition concerns arcs from $X$ to $Y$; regularity in the reverse direction is a separate condition. For $\mu\in(0,1)$ and $p\in(0,1]$, an $n$-vertex undirected graph $G$ is \emph{$(\mu,p)$-upper-uniform} if $|E_G(X,Y)|\le(1+\mu)p|X||Y|$ for every pair of disjoint sets $X,Y\subseteq V(G)$ with $|X|,|Y|\ge\mu n$. For $\xi \in (0, 1)$ and $q \in (0, 1]$, a bipartite graph $H$ with non-empty vertex classes $X, Y$ is \emph{$(\xi, q)$-lower-regular} if $|E_H(X', Y ')| \geq (1 - \xi)q|X'||Y'|$ whenever $X' \subseteq X$ and $Y' \subseteq Y$ satisfy $|X'| \geq \xi|X|$ and $|Y'| \geq \xi|Y|$. We use the same terminology for the pair $(X,Y)$. For disjoint non-empty vertex sets $X,Y$ in an
oriented graph $D$, we call the ordered pair $(X,Y)$
$(\xi,q)$-lower-regular if the same condition holds
with $E_D(X',Y')$ instead of $E_H(X',Y')$.

We write $G\sim G(n,p)$ for the Erd\H{o}s-R\'enyi random graph on $[n]$, in which each edge is present independently with probability $p$.

\section{Structural Tools}
\label{s:properties}
In this section, we introduce the structural tools we require.

\subsection{Building the template graph}

Our algorithm starts by querying the edges of a sparse $n$-vertex undirected graph that we call the \emph{template graph} (its parameters only depend on $\varepsilon$). This graph has at most $n^{4/3}$ edges, which we choose and fix in advance so that, whatever answers we receive, we can use them to select a small set of $O_\varepsilon(n^{2/3})$ vertices that, once we query the induced subtournament, reveal a vertex that controls almost the entire tournament.

\begin{definition}
\label{def:template}
Let $\kappa\in(0,1)$. An undirected graph $G$ on $n$ vertices is a \emph{$\kappa$-template graph} if it has at most $n^{4/3}$ edges and, for every pair of disjoint sets $X,Y\subseteq V(G)$ with
$|X|,|Y|\geq \kappa n$, satisfies
\[
    (1-\kappa){n^{-2/3}}|X||Y|
    \leq |E_G(X,Y)|
    \leq (1+\kappa){n^{-2/3}}|X||Y|.
\]
\end{definition}

\noindent We now show that for any choice of $\kappa\in(0,1)$, such a graph always exists when $n$ is large enough.

\begin{lemma}
\label{template-existence} 
For all $\kappa\in(0,1)$, there exists a $\kappa$-template $n$-vertex graph for sufficiently large $n$.
\end{lemma}

\begin{proof}
Fix $\kappa\in(0,1)$ and choose $G\sim G(n,n^{-2/3})$. For any fixed disjoint sets $X,Y\subseteq[n]$ of size at least $\kappa n$, the random variable $|E_G(X,Y)|$ has expectation $n^{-2/3}|X||Y|$. By a Chernoff bound, the probability that it violates either inequality in Definition~\ref{def:template} is at most $2\exp(-\kappa^2 n^{-2/3}|X||Y|/3)\leq 2\exp(-\kappa^4 n^{4/3}/3)$. There are at most $3^n$ ordered pairs of disjoint sets, since each vertex belongs to $X$, to $Y$, or to neither. Thus, by a union bound, the probability that any such pair violates the required inequalities is at most $2\exp(n\ln 3-\kappa^4 n^{4/3}/3)=o(1)$.

Moreover, $\mathbb{E}[|E(G)|]=\binom{n}{2}n^{-2/3}<n^{4/3}/2$, so Markov's inequality gives $\Pr[|E(G)|>n^{4/3}]<1/2$. By another union bound, with positive probability $G$ has at most $n^{4/3}$ edges and satisfies all the required inequalities, provided $n$ is sufficiently large.
\end{proof}

\subsection{Nice properties of any orientation of the template graph}

Since we fix the template graph before we even query the first edge of the input graph, creating it costs nothing query-wise. We then query its edges and show that any orientation must reveal enough structure to let us partition its vertex set into a constant number of \emph{groups}, with a very small additional leftover group. All but a very small number of pairs of groups have strong enough connectivity that we can choose a \emph{court} (i.e.~some specific subset) of $O_\varepsilon(n^{2/3})$ vertices in each group with the following property: If most arcs (again, directions come from the queried edges) go from the group $i$ to the group $j$, then any sufficiently large subset of the court in $V_i$ dominates almost the entirety of $V_j$ through queried edges of the template graph. Here, ``sufficiently large'' means a fixed fraction of the court. We formalise these properties below, but first introduce
the notion of a \emph{group graph}.

\begin{definition}[Group graph]
    \label{def:group-graph}
    Let $D$ be an oriented graph and let
    $P=(V_0,V_1,\ldots,V_r)$ be a partition of its
    vertices, where $V_1,\ldots,V_r$ are non-empty.
    Fix $\alpha\in(0,1)$ and $p\in(0,1]$.
    The \emph{group graph} $R(D,P)$ is the oriented graph
    on $[r]$ containing the arc $i\to j$ if and only
    if both of the following hold:
    \begin{itemize}
        \item Both ordered pairs $(V_i,V_j)$ and
        $(V_j,V_i)$ are $(\alpha,p)$-regular in $D$.
        \item Either $d_D^+(V_i,V_j)>d_D^+(V_j,V_i)$,
        or $d_D^+(V_i,V_j)=d_D^+(V_j,V_i)$ and $i<j$.
    \end{itemize}
\end{definition}

\begin{lemma}[Four nice properties of template graphs]
\label{lem:nice-properties}
For every $\eta\in(0,1/16]$, there exist
$\alpha,\kappa\in(0,1)$ such that the following holds
for all sufficiently large $n$.
Let $G$ be a $\kappa$-template graph on $n$ vertices.
Every orientation $D$ of $G$ admits a partition
$P=(V_0,V_1,\ldots,V_r)$ and courts
$A_i\subseteq V_i$ for $i\in[r]$ with the following
properties:
\begin{enumerate}
    \item $|V_0|\leq\eta n$, $|V_i|=\ell$ for every
    $i\in[r]$, and
    $\lceil1/\eta\rceil\leq r=O_\eta(1)$.

    \item Let $R \coloneq R(D,P)$ be the group graph with $\alpha$ and $p=n^{-2/3}$
    as in Definition~\ref{def:group-graph}.
    There are at most $\eta r^2$ unordered pairs of
    distinct indices $i,j\in[r]$ such that
    $(i,j) \notin E(R)$ and $(j,i) \notin E(R)$.

    \item The courts have a common size
    $L\geq1/\eta$, where $L=O_\eta(n^{2/3})$.

    \item For every arc $i\to j$ in $R$ and every
    $Z\subseteq A_i$ with $|Z|\geq\eta L$, we have
    $|N_D^+(Z)\cap V_j|\geq(1-\eta)\ell$ (illustrated in Figure~\ref{fig:court-property}).
\end{enumerate}
The partition, group graph, and courts can all be
determined from $D$ alone.
\end{lemma}
\usetikzlibrary{patterns,positioning}

\begin{figure}
    \centering
    \begin{tikzpicture}[
        vertex/.style={
            circle, fill=black,
            minimum size=0.15cm, inner sep=0pt
        }
    ]
    \draw[thick] (0,0) rectangle (4,3);
    \node at (2,3.3) {$V_i$};

    \draw[thick] (0,0) rectangle (4,1.5);
    \node at (1.5,0.75) {$A_i$};

    \draw[thick,fill=black!10] (2.7,0) rectangle (4,1);
    \node at (3.15,0.5) {$Z$};
    \node at (3.4,-0.4) {$|Z|\geq\eta L$};

    \node[vertex] (z1) at (3.65,0.2) {};
    \node[vertex] (z2) at (3.65,0.5) {};
    \node[vertex] (z3) at (3.65,0.8) {};

    \draw[thick] (8,0) rectangle (12,3);
    \node at (10,3.3) {$V_j$};

    \draw[thick,pattern=north east lines]
        (11.5,0) rectangle (12,3);
    \node at (11.75,-0.4) {$\leq\eta|V_j|$};

    \foreach \name/\x/\y in {
        a1/9/1,
        a2/10/1.3,
        a3/9.8/1.9,
        a4/10.5/2.3,
        a5/11/0.5,
        a6/8.9/2.5,
        a7/11.75/0.5,
        a9/11.75/1.5,
        a11/11.75/2.5
    }{
        \node[vertex] (\name) at (\x,\y) {};
    }

    \draw[thick,->] (4.7,2.8) -- (7.3,2.8)
        node[midway,above] {$i\to j \in R$};

    \foreach \source/\target in {z1/a1,z1/a5,z2/a2,z2/a3,z3/a4,z3/a6}{
    \draw[->,shorten >=2pt,shorten <=2pt]
        (\source) -- (\target);
}

    \node at (9.7,-0.4) {$\geq(1-\eta)|V_j|$};
    \end{tikzpicture}

    \caption{The court property for an arc $i\to j$ of $R$.
    Every subset $Z\subseteq A_i$ with $|Z|\geq\eta L$
    sends queried arcs (from the template graph) to at least $(1-\eta)|V_j|$
    vertices of $V_j$. The shaded region contains
    the vertices receiving no arc from $Z$.}
    \label{fig:court-property}
\end{figure}
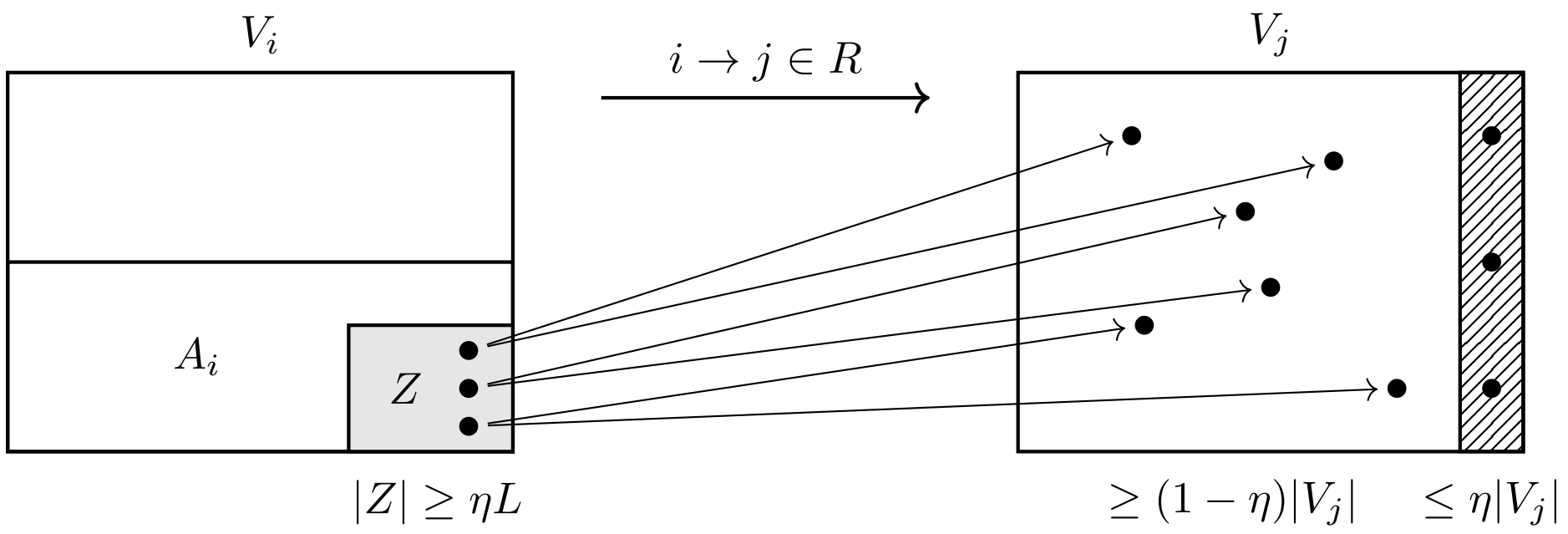

\subsection{Courts hide a powerful vertex} We defer the proof of Lemma~\ref{lem:nice-properties} to Section~\ref{s:structural}. Once we have these properties, we need the final piece of the puzzle: No matter how the remaining pairs of vertices in the union of the courts are oriented, there will always be a vertex that beats (i.e.,~has an outgoing arc to) enough vertices in the right courts for those vertices, together, to dominate almost all of the graph. This gives us our $(1-\varepsilon)$-king. Since each court has size $O_\varepsilon(n^{2/3})$ and there are only a constant number of courts, querying all remaining pairs in their union takes $O_\varepsilon(n^{4/3})$ queries, which keeps us within our total query budget.

\begin{definition}[Active groups]
\label{def:active-groups}
Let $A_1,\ldots,A_r$ be pairwise disjoint courts of common
size $L$, and let $S$ be a tournament on their union
$Q$. Fix $\eta\in(0,1)$.
A group $i\in[r]$ is \emph{active} for a vertex $w\in Q$ if
$|N_S^+(w)\cap A_i|\geq\lceil\eta L\rceil$.
We write $I_S(w)$ for the set of active groups of $w$.
\end{definition}

\begin{lemma}[Powerful vertex hidden inside the courts]
\label{lem:active-courts}
For every $\tau\in(0,1)$, there exists
$\eta\in(0,\min\{\tau,1/16\}]$ such that the following
holds. Let $R$ and $A_1,\ldots,A_r$ be the group
graph and courts from
Lemma~\ref{lem:nice-properties} with $\eta$.
For every tournament $S$ on the union of the courts,
some vertex $w\in V(S)$ satisfies
$|N_R^+(I_S(w))|\geq(1-2\tau)r$.
\end{lemma}

\noindent Again, we defer the proof to Section~\ref{s:structural}. Using Lemma~\ref{lem:nice-properties} and Lemma~\ref{lem:active-courts}, we can now put together the algorithm for obtaining a $(1-\varepsilon)$-king.

\section{Algorithm}
\label{s:algorithm}

Throughout, assume that $n$ is sufficiently large; otherwise, for the finitely many smaller values of $n$, simply query the whole tournament. First, fix $\varepsilon \in (0,1)$. Set $\tau=\varepsilon/8$. Choose $\eta$ as guaranteed
by Lemma~\ref{lem:active-courts} for this value of
$\tau$, and then choose $\alpha,\kappa$ as guaranteed
by Lemma~\ref{lem:nice-properties} for this value of
$\eta$. All these constants depend only on $\varepsilon$. Let $T=([n], \emptyset)$ be the queried graph, initially empty. The algorithm then proceeds in two phases:

\begin{enumerate}
    \item Choose a $\kappa$-template graph $G$ on $[n]$.
    Query each edge of $G$ and add the obtained arc to $T$.
    Denote the resulting orientation of $G$ by $D$.

    \item Use $D$ to find the courts given by
    Lemma~\ref{lem:nice-properties}, and let
    $Q=\bigcup_{i=1}^r A_i$.
    Query the remaining unqueried arcs in the induced subtournament on $Q$ and add the obtained arcs to $T$. Let $S=T[Q]$.
\end{enumerate}

\noindent Once these two steps have been performed, we find a vertex $w\in Q = V(S)$ satisfying
$|N_R^+(I_S(w))|\geq(1-2\tau)r$, and output $w$. We now show that such a vertex always exists, thereby proving the correctness of the algorithm.

\subsection{Proof of correctness and query complexity of the algorithm}

First, we require the following observation.
\begin{observation}\label{obs:tau-eta}
    Let $0\leq\tau\leq1/2$ and $\eta\geq0$. Then  $(1-2\tau)(1-\eta)^2\geq1-2\tau-2\eta$.
\end{observation}
\begin{proof}
     We have $(1 - \eta)^2 = 1 - 2\eta + \eta^2 \geq 1-2\eta$. Since $\tau \leq 1/2$, we have $1 - 2\tau \geq 0$ and hence $(1-2\tau)(1-\eta)^2\geq(1-2\tau)(1-2\eta) = 1 - 2\tau - 2\eta + 4\tau\eta\geq1-2\tau-2\eta$, where the last inequality holds since $\tau, \eta \geq 0$.
\end{proof}

\noindent By Lemma~\ref{lem:active-courts}, applied to $S$, there
exists $w\in Q$ such that
\begin{equation}\label{eq:goodw}
    |N_R^+(I_S(w))|\geq(1-2\tau)r.
\end{equation}

\noindent For each $j\in N_R^+(I_S(w))$, choose an arbitrary $i\in I_S(w)$ with $(i,j)\in E(R)$ and define $\mathrm{Pre}_j(w):=N_S^+(w)\cap A_i$. By Definition~\ref{def:active-groups}, since $i$ is an active group for $w$ we have $|\mathrm{Pre}_j(w)| = |N_S^+(w) \cap A_i| \geq \lceil\eta L\rceil$. By Property~4 of Lemma~\ref{lem:nice-properties},
\begin{equation}\label{eq:size-prej}
    |\mathrm{Pre}_j(w)|\geq\lceil\eta L\rceil \quad\Longrightarrow\quad |N_D^+(\mathrm{Pre}_j(w))\cap V_j|\geq(1-\eta)\ell,
\end{equation}
where $L=|A_i|$ and $\ell=|V_i|$ for all $i\in[r]$. Moreover, $N_D^+(\mathrm{Pre}_j(w)) \cap V_j \subseteq N_T^{++}(w) \cap V_j$, since each vertex on the left is reachable from $w$ through an arc of $S$ followed by an arc of $D$, queried in the second and first batches, respectively (and cannot be $w$ itself since $S,D$ are subgraphs of the same tournament, and $\mathrm{Pre}_j(w) \subseteq N_S^+(w)$). Thus, \eqref{eq:size-prej} implies
\begin{equation}\label{eq:two-step-per-group}
|N_T^{++}(w)\cap V_j|\geq |N_D^+(\mathrm{Pre}_j(w))\cap V_j|\geq(1-\eta)\ell.
\end{equation}
Since the groups $V_j$ are pairwise disjoint, \eqref{eq:two-step-per-group} gives
\begin{equation}\label{eq:nt++}
|N_T^{++}(w)|\geq\sum_{j\in N_R^+(I_S(w))}|N_T^{++}(w)\cap V_j|\geq |N_R^+(I_S(w))|(1-\eta)\ell.
\end{equation}
Thus, by \eqref{eq:nt++},
\begin{align*}
|N_T^{++}(w)|
&\geq |N_R^+(I_S(w))|(1-\eta)\ell \\
&\geq (1-2\tau)(1-\eta)r\ell\tag*{by \eqref{eq:goodw}}\\
&\geq (1-2\tau)(1-\eta)^2n\tag*{since $r\ell=n-|V_0|\geq(1-\eta)n$ by Property~1 of Lemma~\ref{lem:nice-properties}}\\
&\geq (1-2\tau-2\eta)n \tag*{by Observation~\ref{obs:tau-eta}}\\
&\geq (1-\varepsilon)n,
\end{align*}
where the last inequality follows because
$\eta\leq\tau=\varepsilon/8$. Lastly, the total number of queries is clearly $O_\varepsilon(n^{4/3})$ as $|E(G)| \leq n^{4/3}$ and $|Q|=O_\varepsilon(n^{2/3})$, which is guaranteed by Property~3 of Lemma~\ref{lem:nice-properties}.

\subsection{An extension to two tournaments}
Afek, Gafni and Linial~\cite{AGL19} showed that, for any two tournaments on the same vertex set, there is a vertex that reaches every other vertex by an arc in either tournament, or by an arc in the first followed by an arc in the second. Our algorithm can easily be generalised to efficiently find an approximate version of such a vertex.




\begin{corollary}\label{cor:two-tournaments}
For every fixed $\varepsilon\in(0,1)$, there is a deterministic algorithm that, given two tournaments $T_1,T_2$ on the same $n$-vertex set, finds a vertex $w$ from which at least $(1-\varepsilon)(n-1)$ other vertices are reachable by an arc in either tournament, or by an arc in $T_1$ followed by an arc in $T_2$, in $O_\varepsilon(n^{4/3})$ queries.
\end{corollary}

\begin{proof}
We use the same algorithm with $\varepsilon/2$ instead of $\varepsilon$, with one change: We query the template graph edges in $T_2$, and all pairs within the union $Q$ of the courts in $T_1$. The orientation of the template graph in $T_2$ gives us the partition, the directed graph $R$ and the courts. Since Lemma~\ref{lem:active-courts} holds for any tournament on $Q$, we may apply it to $T_1[Q]$. Thus, $w$ has arcs into the courts through arcs in $T_1$, while Property~4 of Lemma~\ref{lem:nice-properties} gives the arcs from the courts to the target groups in $T_2$.

The same counting argument shows that at least $(1-\varepsilon/2)n$ vertices are reached by an arc in $T_1$ followed by an arc in $T_2$. Here, $w$ itself may be counted, since the two arcs belong to different tournaments. Excluding it leaves at least $(1-\varepsilon/2)n-1\ge(1-\varepsilon)(n-1)$ other vertices, as required.
\end{proof}

\section{Proofs of the Structural Tools}
\label{s:structural}
This section is devoted to the proofs of Lemma~\ref{lem:nice-properties} and Lemma~\ref{lem:active-courts}.

\subsection{Proof of Lemma~\ref{lem:nice-properties}: four nice properties of template graphs}

We first state the following extension of Szemer\'edi's regularity lemma due to Allen et al.~\cite{AKMP14} (also see~\cite{kohayakawa1997szemeredi, kohayakawa2003szemeredi}).

\begin{theorem}
[{\cite[Theorem~2.1]{AKMP14}}]
\label{regularity-lemma}
    For all $\tau > 0$ and positive integer $m$, there exist $\mu \in (0,1)$ and $M \geq m$ such that, for any $p \in (0,1]$ and any orientation $H'$ of a sufficiently large $n$-vertex $(\mu,p)$-upper-uniform undirected graph $H$, there exists a partition of $V(H')$ into $k+1$ groups $V_0 \sqcup V_1 \sqcup V_2 \sqcup \cdots \sqcup V_k$ with $m \leq k \leq M$ such that $|V_1|=|V_2|= \cdots = |V_k|$, among all distinct $i, j \in [k]$, at most $\tau k^2$ ordered pairs $(V_i, V_j)$ are not $(\tau, p)$-regular, and $|V_0| \leq \tau n$.
\end{theorem}

\noindent We also need the following inheritance result for lower-regularity, which is~\cite[Theorem~3.6]{GKRS07} with $\beta=1/2$.

\begin{theorem}[{\cite[Theorem~3.6]{GKRS07}}]
\label{lower-regularity-inheritance}
For every $\chi\in(0,1)$, there exist constants $\xi_0,K>0$ such that the following holds. Let $0<\xi\leq\xi_0$, let $q\in(0,1)$, and let $H$ be a $(\xi,q)$-lower-regular bipartite graph with $V(H)=X\sqcup Y$. For every integer $L$ with $K/q\leq L\leq|X|$, there are at least
$$
(1-2^{-L})\binom{|X|}{L}
$$
subsets $A\subseteq X$ of size $L$ such that $(A,Y)$ is $(\chi,q)$-lower-regular.
\end{theorem}

\noindent We can now prove that each property of Lemma~\ref{lem:nice-properties} holds for any choice of $\eta \in (0,1/16]$. 

\begin{proof}[Proof of Lemma~\ref{lem:nice-properties}]
Fix $\eta\in(0,1/16]$ and let $p=n^{-2/3}$.
Apply Theorem~\ref{lower-regularity-inheritance} with $\chi=\eta/2$ to obtain $\xi_0,K>0$, and choose $\alpha=\min\{\eta,\xi_0/2\}.$
Apply Theorem~\ref{regularity-lemma} with $\tau=\alpha$ and $m=\lceil1/\eta\rceil$ to obtain $\mu,M$. Finally, set $\kappa=\min\{\mu/2,1/(2M)\}.$
All these constants depend only on $\eta$.

\paragraph{Property 1.} Let $G$ be any $\kappa$-template graph for sufficiently large $n$, and fix an orientation $D$ of $G$. The inequalities of Definition~\ref{def:template} imply that $G$ is $(\mu,p)$-upper-uniform. Theorem~\ref{regularity-lemma} therefore
gives a partition $V(D)=V_0\sqcup V_1\sqcup\cdots\sqcup V_r$ with $m\leq r\leq M$, $|V_0|\leq\alpha n\leq\eta n$, and $|V_i|=\ell$ for all $i\in[r]$. This proves Property 1. Moreover,
$
\ell=\frac{n-|V_0|}{r}\geq\frac{n}{2M}\geq\kappa n,
$
so the inequalities of Definition~\ref{def:template} apply for any pair of distinct groups $V_i,V_j$.

\paragraph{Property 2.} Now, construct $R$ as in the statement. Every missing unordered pair has at least one direction that is not $(\alpha,p)$-regular. There are at most $\alpha r^2$ such ordered pairs, so at most $\eta r^2$ unordered pairs do not have an arc in $R$. This proves Property 2.

\paragraph{Property 3.} We now choose the courts. For each arc $(i,j)\in E(R)$, let $H_{ij}$ be the undirected bipartite graph on $V_i,V_j$ whose edges are precisely the edges corresponding to the arcs of $D$ directed from $V_i$ to $V_j$. Since this is the majority direction, by definition, and $\kappa\leq 1/(2M) \leq 1/32 < 1/4$, its density is at least
$
\rho_D(V_i,V_j)\geq(1-\kappa)p/2
\geq 3p/8.
$
For any $X\subseteq V_i$ and $Y\subseteq V_j$ with
$|X|,|Y|\geq2\alpha\ell$, regularity gives
$
\frac{|E_{H_{ij}}(X,Y)|}{|X||Y|}
\geq 3p/8-\alpha p
\geq p/4.
$

Thus $H_{ij}$ is $(2\alpha,p/4)$-lower-regular, so set $L=\lceil4K/p\rceil$. For sufficiently large $n$,
we have $1/\eta\leq L\leq\ell$.
Choose each $A_i$ independently and uniformly among
all subsets of size $L$ of $V_i$.
Since $2\alpha\leq\xi_0$, Theorem~\ref{lower-regularity-inheritance} shows that, for each $(i,j)\in E(R)$, the pair $(A_i,V_j)$ is $(\chi,p/4)$-lower-regular except
with probability at most $2^{-L}$. A union bound over all arcs of $R$ gives that the probability that at least one pair $(A_i,V_j)$ is not $(\chi,p/4)$-lower-regular over all arcs $(i,j)\in E(R)$ is at most $r^2\,2^{-L}\leq M^2\,2^{-L}<1$ for sufficiently large $n$. We can therefore fix
courts for which all these pairs are lower-regular, which proves Property 3. 

\paragraph{Property 4.}
Fix $(i,j)\in E(R)$ and $Z\subseteq A_i$ with $|Z|\geq\eta L$. Let $B=V_j\setminus N_D^+(Z)$. Suppose for a moment that $|B|\geq\eta\ell$. Since $\chi=\eta/2$, we have $|Z|\geq\chi|A_i|$ and $|B|\geq\chi|V_j|$. The $(\chi,p/4)$-lower-regularity of $(A_i,V_j)$ therefore gives
$
d^+_D(Z,B)\geq(1-\chi)\frac p4|Z||B|>0,
$
contradicting the definition of $B$ as $Z$ has no arc into $B$. Thus
$|B|<\eta\ell$, and consequently
$
|N_D^+(Z)\cap V_j|\geq(1-\eta)\ell.$\end{proof}

\subsection{Proof of Lemma~\ref{lem:active-courts}: powerful vertex hidden inside the courts}

We first use the following combinatorial result of Afek et al.~\cite{AGL19}. They prove that, given two tournaments on the same vertex set, there is always a vertex that reaches every other vertex by an arc in either tournament, or by an arc in the first
tournament followed by an arc in the second. We include a proof for completeness.

\begin{lemma}[{\cite[Theorem~1]{AGL19}}]
\label{two-tournaments}
Let $T_1,T_2$ be tournaments on $[n]$.
Then there exists a vertex $v\in[n]$ such that
$$
\{v\}\cup N^+_{T_1}(v)\cup N^+_{T_2}(v)
\cup N^+_{T_2}\bigl(N^+_{T_1}(v)\bigr)=[n].
$$
\end{lemma}
\begin{proof}
Here, say that $x$ reaches $y$ if $x=y$, there is an arc
$x\to y$ in either tournament, or there is a path
$x\to z\to y$ whose first arc belongs to $T_1$ and
second arc belongs to $T_2$.

The claim for $n \leq 3$ can be easily verified. Suppose now, for the sake of contradiction, that the statement is false in general for $n > 3$, and choose a smallest counterexample on a pair of tournaments $A$ and $B$ of size $n$. By minimality,
for every $v\in [n]$, there is a vertex $x_v\neq v$
that reaches every vertex in $A[[n]\setminus\{v\}]$ and $B[[n]\setminus\{v\}]$; thus $x_v$ reaches
every vertex other than $v$ in $A$ and $B$, and it cannot reach $v$ (since otherwise
it would reach all vertices). Also observe that the vertices $x_v$ are distinct: Indeed, if $w=x_u=x_v$ for
distinct $u,v$, then $w$ would reach $u$ because $w=x_v$ (i.e.~$w$ reaches all vertices in $A[[n]\setminus\{v\}]$ and $B[[n]\setminus\{v\}]$) and would not reach $u$ because $w=x_u$. Hence $v\mapsto x_v$ is a
permutation of $[n]$.

Since $x_v$ cannot reach $v$, the vertex $v$ beats
$x_v$ in $B$ and also beats in $B$ every out-neighbour
of $x_v$ in $A$. Therefore, we have
$d^+_{B}(v)>d^+_{A}(x_v)$.
Summing over $v\in [n]$ gives
$\binom{n}{2}
=\sum_{v\in [n]}d^+_{B}(v)
>\sum_{v\in [n]}d^+_{A}(x_v)
=\binom{n}{2}$,
which is impossible.
\end{proof}

\noindent We can now prove that Lemma~\ref{lem:active-courts} holds for any $\tau \in (0,1)$. An informal sketch of the proof is as follows: towards a contradiction, assume that every vertex has
many groups outside both its active groups and their
out-neighbourhood in $R$. We sample a small number
of courts and one representative from each. We show that, with positive probability, every pair
of sampled groups has an arc in $R$, every
representative activates its own court, and every
sampled arc $x_a\to x_b$ in $S$ ends in a court
active for $x_a$. The last condition is likely because
an inactive court contains very few out-neighbours
of $x_a$. Our assumption also ensures that the sample
is likely to contain, for every representative, a
group outside both its active groups and their
out-neighbourhood in $R$. Now suppose we fix a sample satisfying all these conditions.
The representatives induce a tournament in $S$,
and their court indices induce a tournament in $R$.
Identify each representative with its court index.
A direct arc in the first tournament ends in an
active group. A direct arc in the second tournament
ends in an out-neighbour of an active group,
because each representative activates its own court.
The same holds for a path consisting of an arc
in the first tournament followed by an arc in the
second. Thus, none of these routes reaches the
sampled group outside both sets, contradicting
Lemma~\ref{two-tournaments}. Hence, some vertex has few groups outside its active
groups and their out-neighbourhood. Finally, any two
active groups lying outside that out-neighbourhood
have no arc between them in $R$. Since $R$ has few
such pairs, there can be only a small number of these
groups.

\begin{proof}[Proof of Lemma~\ref{lem:active-courts}]
Fix $\tau\in(0,1)$. Choose an integer $s\geq2$ with
$se^{-\tau(s-1)}\leq1/4$, and choose
$
0<\eta\leq\min\{1/16,\tau^2/8,1/(8s^2)\}.
$
Then $r\geq\max\{s,2/\tau\}$ and $\eta L\geq1$.
For each $w\in Q$, define
$
B(w)=[r]\setminus\bigl(I_S(w)\cup N_R^+(I_S(w))\bigr).
$
We show that there exists a vertex $w$ such that $|B(w)| < \tau r$.  For the sake of contradiction, suppose that for all $w \in Q$, we have $|B(w)| \geq \tau r$. We will find an ordered tuple of $s$ groups and
representatives that contradicts Lemma~\ref{two-tournaments}. Choose distinct group indices $i_1,\ldots,i_s$ uniformly,
then independently choose uniform representatives
$x_a\in A_{i_a}$. Consider three bad scenarios:
\begin{enumerate}
    \item Some $x_a$ has no selected group in $B(x_a)$.
    Conditional on $i_a,x_a$, this has probability zero
    if $i_a\in B(x_a)$, and otherwise at most
    $(1-|B(x_a)|/(r-1))^{s-1}\leq e^{-\tau(s-1)}$.
    A union bound gives probability at most
    $se^{-\tau(s-1)}\leq1/4$.

    \item Some selected pair is missing in $R$.
    Its probability is at most
    $
    \binom{s}{2}\frac{\eta r^2}{\binom r2}
    \leq2\eta s^2\leq1/4.
    $

    \item Some $i_a\notin I_S(x_a)$, or some
    $(x_a,x_b) \in E(S)$ has $i_b\notin I_S(x_a)$.
    Put $h=\lceil\eta L\rceil$.
    At most $2h-1\leq3\eta L$ vertices of a court
    have out-degree within the court below $h$: if there are
    $t>0$ such vertices, then
    $\binom t2\leq t(h-1)$.
    Thus we have $i \not\in I_S(x)$ with probability
    at most $3\eta$ per representative $x$ of group $i$.
    For distinct $a,b$, condition on the indices and
    $x_a$. If $i_b\notin I_S(x_a)$, fewer than $\eta L$
    choices of $x_b$ are dominated by $x_a$.
    A union bound therefore gives probability at most
    $
    3\eta s+\eta s(s-1)\leq2\eta s^2\leq1/4.
    $
\end{enumerate}

\noindent A union bound says that the total probability that at least one bad event happens is at most $3/4$, so we can fix some sample avoiding all three scenarios. Define tournaments $T_1$ and $T_2$ on $[s]$ by
$$
(a,b) \in E(T_1)\iff (x_a,x_b) \in E(S),
\qquad
(a,b) \in E(T_2)\iff (i_a,i_b) \in E(R).
$$
Note that $T_2$ is a tournament because the second bad scenario does not occur. Take any $a\in[s]$. Since the first bad scenario does not occur, there exists $b$ such that $i_b\in B(x_a)$. The third bad scenario does not occur either, so $i_a\in I_S(x_a)$, which gives $a\ne b$.

Neither $(a,b) \in E(T_1)$ nor $(a,b) \in E(T_2)$ is possible: the first would put $i_b$ in $I_S(x_a)$, and the second would put it in $N_R^+(I_S(x_a))$, both contradicting
$i_b\in B(x_a)$. Now suppose some $c$ satisfies $(a,c) \in E(T_1)$ and $(c,b) \in E(T_2)$. Since the third bad scenario does not occur, $i_c\in I_S(x_a)$. But $(i_c,i_b) \in E(R)$, so $i_b\in N_R^+(I_S(x_a))$, again a contradiction. Thus every $a$ has a vertex $b$ that none of the three
allowed paths reaches, which directly contradicts
Lemma~\ref{two-tournaments}.

We have therefore found $w\in Q$ with $|B(w)|<\tau r$.
Let $J=I_S(w)\setminus N_R^+(I_S(w))$.
Every pair in $J$ is missing in $R$, since an arc
between two members would put its target in
$N_R^+(I_S(w))$. Hence
$
\binom{|J|}{2}\leq\eta r^2,
$ and $
|J|\leq\sqrt{2\eta}\,r+1
\leq\frac{\tau r}{2}+1
\leq\tau r,
$
using $\eta\leq\tau^2/8$ and $r\geq2/\tau$.
Finally,
$[r]\setminus N_R^+(I_S(w))=B(w)\sqcup J$, so
$
|N_R^+(I_S(w))|
=r-|B(w)|-|J|
\geq(1-2\tau)r,
$
as required.
\end{proof}

\section{Concluding Remarks}
\label{s:conclusion}

We have shown that, for every fixed $\varepsilon\in(0,1)$, a $(1-\varepsilon)$-king can be found deterministically using $O_\varepsilon(n^{4/3})$ queries. This matches the known lower bound~\cite{AFHN15, BJRS22} and answers the main open question of Lachish, Reidl and Trehan~\cite{LRT22} for every fixed fraction below one. The algorithm uses only two batches: we first query the template graph, then use its orientation to partition the vertices into groups, find a court within each group, and query the subtournament induced by the union of the courts.

Naturally, the main question remains whether a full king can be found with $O(n^{4/3})$ deterministic queries. One difficulty is that finding an approximate king does not give a cheap way to check whether it is already a full king. In general, even deciding whether a given vertex $w$ is a king requires $\Omega(n^2)$ queries~\cite{BJRS22}.

\subsection*{A potential route towards a 1-king}

Fix $w\in Q$. Call a vertex $v\ne w$ \emph{certified} if either the arc $w\to v$ has been queried, or there exists $z\in Q$ such that both arcs $w\to z$ and $z\to v$ have been queried. Let $U$ be the set of vertices other than $w$ that are not certified. We show that reducing $|U|$ to $O_\varepsilon(n^{2/3})$ would suffice to find a full king using $O_\varepsilon(n^{4/3})$ additional queries. If $|U|=O_\varepsilon(n^{2/3})$, we can query all remaining pairs within $Q\cup U$ using $O_\varepsilon(n^{4/3})$ additional queries. If $w$ is a king of this subtournament, it is a king of the whole tournament. Otherwise, let $B$ be the non-empty set of other vertices that $w$ does not control within this subtournament. Every vertex in $B$ beats $w$ and all of its out-neighbours within $Q\cup U$. It therefore reaches every vertex outside $B$ within two steps, using $w$ or one of those out-neighbours as an intermediary. Consequently, a king of the subtournament induced by $B$ is a full king, and all edges needed to find it have already been queried.

The obstacle is that our present argument leaves up to $\varepsilon n$ vertices uncertified. For fixed $\varepsilon$, this can be much larger than $n^{2/3}$. Improving the dependence on $\varepsilon$ may help here, but the constants in our bound depend on $\varepsilon$ through the regularity and inheritance results. We therefore cannot simply let $\varepsilon$ decrease with $n$ and retain the same query bound. On the lower-bound side, any bound stronger than $\Omega(n^{4/3})$ for finding a full king must distinguish reaching every vertex from reaching all but an arbitrarily small fixed fraction.

Finally, our algorithm is not time-efficient: the existence of the template graph, the partition and the courts, all use existential arguments, which is fine for us since computation is free in the query model. Can one design an $O_\varepsilon(n^{4/3})$-\emph{time} algorithm for finding a $(1-\varepsilon)$-king?

\paragraph{AI methodology.}
The algorithmic approach used in this paper was developed by the authors. ChatGPT Astra was used to identify Lemma~\ref{two-tournaments} and Theorem~\ref{lower-regularity-inheritance}, which are used in the proofs of Lemma~\ref{lem:active-courts} and Properties 3 and 4 of Lemma~\ref{lem:nice-properties}, respectively. It was also used for mathematical proofreading, choosing constants, and polishing the exposition. All AI-assisted contributions incorporated into the paper were verified by the authors, who take full responsibility for its content.

\bibliography{bib}

\end{document}